\documentclass[10pt,leqno]{amsart}

\usepackage[utf8]{inputenc}
\usepackage[T1]{fontenc}

\usepackage{amsmath,amssymb,amsthm,amsfonts}
\usepackage{graphicx}
\usepackage{xcolor}
\usepackage{booktabs}
\usepackage{array}
\usepackage{makecell}
\usepackage{multirow}
\usepackage{rotating}
\usepackage{float}
\usepackage{textcomp}
\usepackage{url}
\usepackage{verbatim}
\usepackage{multicol}
\usepackage{microtype}
\usepackage{tikz}
\usetikzlibrary{patterns}
\usepackage{pgfplots}
\usepackage{pgfplotstable}
\pgfplotsset{compat=1.7}
\usepackage{subfig} % ok with amsart
\usepackage{algorithm}
\usepackage{algorithmic}
\usepackage{csquotes}
\usepackage{cite}           % optional with amsart
\usepackage{epstopdf}       % allow .eps with pdflatex (+ -shell-escape)
\usepackage[strings]{underscore} % make _ safe in normal text/strings
\usepackage[hidelinks]{hyperref} % load hyperref LAST

\numberwithin{equation}{section}

\begin{document}

% ---------- Title & Authors (amsart style) ----------
\title[Multi-Omics Interactions for Lung Cancer Targets]{Identifying Multi-Omics Interactions for Lung Cancer Drug Target Discovery Using Kernel Machine Regression}

\author[Ahmed]{Md.~Imtyaz Ahmed}
\address{Department of Information and Communication Technology, Mawlana Bhashani Science and Technology University, Santosh, Tangail, 1902, Dhaka, Bangladesh}
\email{imtyazit17017@gmail.com}

\author[Hossain]{Md.~Delwar Hossain}
\address{Department of Information and Communication Technology, Mawlana Bhashani Science and Technology University, Santosh, Tangail, 1902, Dhaka, Bangladesh}
\email{delwarit14@gmail.com}

\author[Rahman]{Md.~Mostafizer Rahman}
\address{Department of Computer Science, Tulane University, New Orleans, LA, USA}
\email{mrahman9@tulane.edu}
\email{mostafiz26@gmail.com}

\author[Habib]{Md.~Ahsan Habib}
\address{Department of Information and Communication Technology, Mawlana Bhashani Science and Technology University, Santosh, Tangail, 1902, Dhaka, Bangladesh}
\email{mahabib@mbstu.ac.bd}

\author[Rashid]{Md.~Mamunur Rashid}
\address{Bioinformatics Institute (BII), Agency for Science, Technology and Research (A*STAR), 138632, Singapore}
\email{mamunur\_rashid@bii.a-star.edu.sg}

\author[Reza]{Md.~Selim Reza}
\address{Tulane Center for Biomedical Informatics and Genomics, Deming Department of Medicine, Tulane University, New Orleans, LA 70112, USA}
\email{mreza@tulane.edu}

\author[Alam]{Md.~Ashad Alam}
\address{Ochsner Center for Outcomes Research, Ochsner Research, New Orleans, LA 70121, USA}
\email{mdashad.alam@ochsner.org}

% ---------- Abstract & Keywords (BEFORE \maketitle in amsart) ----------
\begin{abstract}
Cancer exhibits diverse and complex phenotypes driven by multifaceted molecular interactions. Recent biomedical research has emphasized the comprehensive study of such diseases by integrating multi-omics datasets (genome, proteome, transcriptome, epigenome). This approach provides an efficient method for identifying genetic variants associated with cancer and offers a deeper understanding of how the disease develops and spreads. However, it is challenging to comprehend complex interactions among the features of multi-omics datasets compared to single omics. In this paper, we analyze lung cancer multi-omics datasets from The Cancer Genome Atlas (TCGA). Using four statistical methods—LIMMA, the t-test, canonical correlation analysis (CCA), and the Wilcoxon test—we identified differentially expressed genes across gene expression, DNA methylation, and miRNA expression data. We then integrated these multi-omics data using the kernel machine regression (KMR) approach. Our findings reveal significant interactions among the three omics: gene expression, miRNA expression, and DNA methylation in lung cancer. From our data analysis, we identified 38 genes significantly associated with lung cancer. Among these, eight genes of highest ranking (PDGFRB, PDGFRA, SNAI1, ID1, FGF11, TNXB, ITGB1, ZIC1) were highlighted by rigorous statistical analysis. Furthermore, in silico studies identified three top-ranked potential candidate drugs (Selinexor, Orapred, and Capmatinib) that could play a crucial role in the treatment of lung cancer. These proposed drugs are also supported by the findings of other independent studies, which underscore their potential efficacy in the fight against lung cancer.
\end{abstract}


\maketitle

\section{Introduction}\label{intro}
In recent years, advances in biomedical technology have led to the accumulation of vast amounts of multi-omics data, revolutionizing disease identification with a holistic approach. Integrating multivariate data aims to uncover intricate interactions among various molecular characteristics, particularly in complex diseases such as lung cancer. Despite the historical focus of the pharmaceutical industry on developing broad-spectrum medicines, personalized treatments tailored to individual patients often yield superior outcomes. Consequently, biomedical researchers are increasingly focused on identifying significant genes linked to complex diseases. However, state-of-the-art methods often fail to optimize the outcomes for these conditions. To address these challenges, the scientific community embraces precision medicine, focusing on personalized treatments driven by comprehensive omics data for diseases such as cancer and schizophrenia \cite{1}. Despite these advancements, the integration of multi-omics data to pinpoint biomarkers for complex diseases presents a formidable challenge. This study aims to use kernel machine regression (KMR) to investigate and uncover critical multi-omics interactions that advance drug target discovery in lung cancer.\\

In state-of-the-art work, omics data analysis can be categorized into two approaches: single-view and dual-view datasets. In single-view-based data analysis, modern high-throughput techniques, such as deep sequencing, generate large volumes of molecular data \cite{2}. This study, for example, includes parameters such as DNA genome sequences \cite{3}, RNA expression levels \cite{4,5}, and DNA methylation patterns \cite{6}. Each type of data is referred to as an ``omic'', including genomics, transcriptomics, and methylomics. Although single-omic approaches can identify biomarkers associated with specific exposures, they often capture only a small subset of biomarkers linked to complex diseases. This limitation hinders their ability to fully elucidate changes in key biological pathways, making them insufficient for a comprehensive understanding of diseases such as lung cancer or prostate cancer \cite{7,8}.
Although it is feasible to conduct a single-omics study for each complex disease, this approach may overlook significant insights. In contrast, integrative omics approaches, although more resource intensive, are widely recognized as valuable tools for acquiring deeper insights into complex diseases \cite{7,9}. Analyzing integrated risk factors and understanding intricate relationships between multiple omics data remain challenging. To fully understand human health and disease, it is essential to interpret the molecular complexity and diversity at various levels, such as the genome, epigenome, proteome, transcriptome, and metabolome \cite{10}. Consequently, employing multi-omics-based data analysis is crucial for cancer detection and drug development. The integration of multi-omics data in cancer detection allows a comprehensive analysis of diverse omics data types, offering a global view of the biological system and providing insights into the relationships between different layers of data \cite{11,12,57}.\\

\begin{figure}
  \centering
  \includegraphics[width=.9\columnwidth]{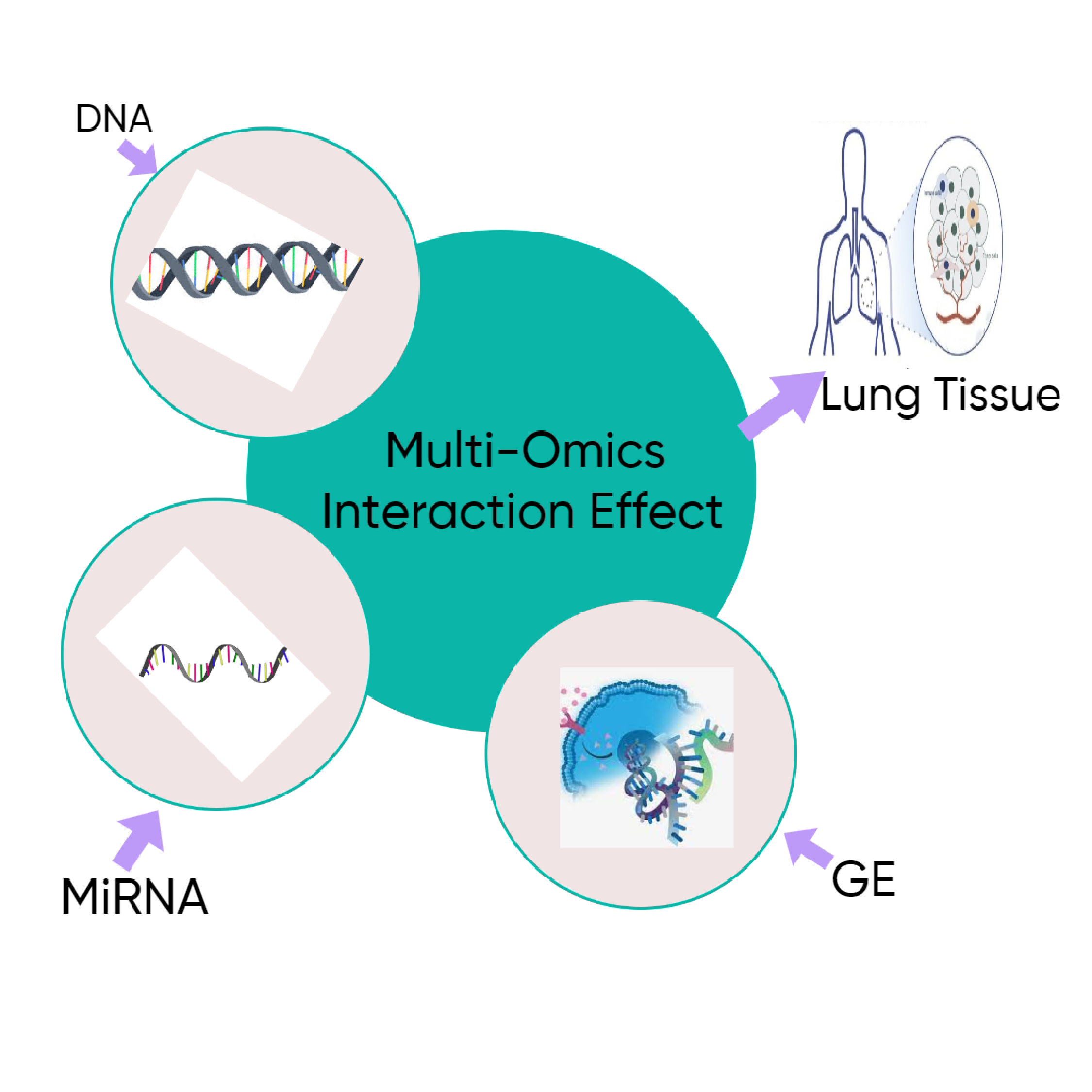}
  \caption{\textbf{The impact of multi-omics data interaction on human lung tissue.} The figure displays the gene expression (GE), micro ribonucleic acid (miRNA), and deoxyribonucleic acid methylations (DNA) data for human lung tissue, and highlights the potential interactions between these omics layers.}
  \label{FIG:1}
\end{figure}

Lung cancer is responsible for the highest number of cancer-related deaths, accounting for almost 25\% of all cancer deaths and is the second most commonly diagnosed cancer globally. Each year, more people pass away from lung cancer than combined breast, colon, and prostate cancers \cite{13}. Lung cancer presents a wide range of symptoms and indications depending on its anatomical development, as it can occur at several points throughout the bronchial tree. A series of genetic and epigenetic alterations are believed to be responsible for transforming a normal lung phenotype into a malignant one, which then proliferates into invasive cancer through clonal expansion. Identifying and characterizing these molecular changes is essential for effective disease prevention, early diagnosis, and treatment \cite{14,58}. However, early-stage diagnosis of lung cancer using multi-omics data analysis remains a significant challenge.
To address this challenge, this study focuses on leveraging KMR to explore and identify critical multi-omics interactions relevant to lung cancer drug target discovery. By integrating data across multiple omics layers, KMR can capture the complex relationships that single-omics approaches may miss.\\
In this paper, we applied the kernel machine regression (KMR) method for integrating a multi-omics dataset of lung cancer data. Linear data integration approaches are extensively used and validated strategies for gaining a more comprehensive understanding of biological processes in complicated diseases. However, traditional methods for data analysis, such as linear techniques, have limitations in dealing with non-linear data structures and multi-modal distributions, resulting in poor performance \cite{15,16}. To address this issue, non-linear integrated techniques such as kernel-based machines have become essential for analyzing multi-omics datasets \cite{17,18}. Positive-definite kernel-based machine approaches have proven effective in resolving the non-linearity issue \cite{19}. Statistical machine learning approaches, such as kernel-based methods, provide valuable information on the relationship between genetic markers and disease states, allowing exploration of a wide range of genetic variants associated with complex traits \cite{20,21}. These methods can help with the efficient integration of multi-omics data. The main focus of this paper is on using Kernel Machine Regression (KMR) to analyze multi-omics interactions for drug repositioning in lung cancer. By integrating gene expression, DNA methylation, and miRNA expression data, the study aims to identify significant interacted genes involved in lung cancer development. This approach provides a more comprehensive understanding of the biological mechanisms and helps in the discovery of drug targets for lung cancer. To verify our results, we performed protein association networks analysis, gene-miRNA-methylation network interaction analysis, molecular coupling analysis, and 2D chemical interaction studies.\\

\section{Preliminaries}\label{prelim}
In the literature, numerous methods have been proposed to evaluate multi-omics datasets, with linear approaches being the most diverse. Popular linear methods for multi-omics following identification analysis include canonical correlation analysis, partial least squares, and multi-omics factor analysis \cite{22,23,24,25}.\\
Liu et al.~(2007) conducted a preliminary study using a single-modal sample to evaluate the effects of the genetic route through a kernel machine technique \cite{26}. In another study, Li and Cui (2012) proposed a machine-based kernel strategy for gene-gene interconnections, where they used each gene as a testing platform and proposed a kernel machine approach to identify various factor relationships using a smoothing spline-ANOVA technique \cite{26}. However, these strategies mostly use single or coupled datasets, which limits their ability to fully capture the complex interplay between multiple omics data sets.\\
Kernel-based techniques are useful for studying how a wide range of genetic variations are related to complex phenotypes and disease states \cite{20,27,28}. For investigating gene-gene co-association, linear, kernel, and robust canonical correlation techniques have been employed \cite{29,30}. Nonlinear kernel-based multi-omics data integration models provide a more comprehensive perspective by combining several data sources and revealing interactions between them. These models are particularly helpful for studying the diverse range of genetic influences connected to intricate phenotypes and disease states \cite{31,32,33}. It is increasingly challenging to identify marginal, interactional, and composite effects in multi-omics datasets.\\
Furthermore, Ge et al.~(2015) suggested using kernel machines to identify the effects of relationships between multidimensional data sets \cite{27}. Another more comprehensive model, which accounts for both genetic and non-genetic components as well as their interactions, was introduced by Guo et al.~(2014) \cite{29,34}. N.~Zhao et al.~(2015) utilized a semi-parametric kernel machine regression framework and introduced the microbiome regression-based kernel association test (MiRKAT) to effectively recover results from single-omics datasets and microbiome profiles \cite{35}. Composite kernel machines and Bayesian variable correlation KMR methods have also been suggested for genome-wide association research \cite{36,37}.
Alam et al.~established a kernel machine technique to identify higher-order interactions in three different datasets and applied it to study schizophrenia with a continuous characteristic \cite{1,57}. In that study, they proposed the Generalized Kernel Machine Approach for Higher-Order Composite Effects (GKMAHCE) to identify composite impacts in multiview biomedical data sources. In research on adolescent brain development and osteoporosis, Alam et al.~used the GKMAHCE technique to analyze synthetic and real multiview sets of data \cite{10,59}. The GKMAHCE method is a generalized semi-parametric method that includes marginal and integrated Hadamard products of characteristics from various points of view of the data, using a mixed-effect linear model \cite{1}. In another study, Jie Feng et al.~used the principal component analysis (PCA) of the kernel to perform gene set enrichment analysis and obtain differential expression of certain genes among different subtypes of lung cancer \cite{38}.

\section{Methodology}\label{method}
\subsection{Data sources}
The dataset on lung cancer analyzed in this article was sourced from the Multi-Omics Cancer Benchmark TCGA Preprocessed Data \cite{41}. \newline\\
\textbf{Lung Squamous Cell Carcinoma (LUSC):} The Cancer Genome Atlas (TCGA), the largest collection of its kind in the USA, gathers and analyzes tumor samples from over 11{,}000 cancer patients. This study measures various aspects of these samples, including tissue genome sequence, copy number variation (CNV), gene expression, microRNA (miRNA) expression, and DNA methylation. In addition, it includes biological and medical data such as racial group, tumor grade, recurrence, and therapeutic response. For this research, we used preprocessed TCGA multi-omics data for LUSC, consisting of gene expression, DNA methylation, and miRNA expression data. Using packages (TCGAbiolinks, EDASeq) of R, we obtained 344 samples with gene expression data, miRNA sequencing (miRNA-Seq) data, DNA methylation data and clinical information in the LUSC data set \cite{41}.\\

\subsection{Material and methods}
\subsubsection{Differential Expression Study}\ \\
\textbf{LIMMA:} Limma is a software package for R/Bioconductor that allows the evaluation of gene expression data from microarray and RNA-Seq experimental analysis. It is designed to handle complex experimental designs and addresses the issue of small sample sizes by incorporating information-borrowing techniques. Limma effectively combines multiple statistical principles to facilitate large-scale expression studies. It provides a unique method for relating new expression data sets to previous experiments, considering factors such as fold change and directional changes for each gene in earlier studies. The package also includes a statistical approach to fitting global covariance models to estimate gene correlations and relatedness between differentially expressed profiles resulting from difference comparisons \cite{40}.\\
\textbf{T-test:}
The t-test is a widely used statistical method to identify genes with up-regulated genes. In replicated experiments, error variance for each gene can be estimated from log ratios and a standard t-test can be performed to identify genes significantly differentially expressed. Unlike other methods, the t-test considers one gene at a time, making it immune to heterogeneity in variance between genes. However, due to the small sample sizes in the number of RNA samples measured for each condition, the t-test may have limited statistical power \cite{41}.
Let the differences between the individual pairs be \(x_i\) and \(y_i\) in individual \(i\) such that \( d_i = x_i - y_i \) and
\begin{equation}
T = \frac{ \sqrt{n} (\bar{d} - \mu_d)}{S_d} \sim t_{n - 1},
\end{equation}
where \(\bar{d}\) represents the average of the differences in the sample, while \( S_d \) denotes the standard deviation of the sample differences.\\
\textbf{Wilcoxon test:} The Wilcoxon test is a non-parametric statistical method used to compare two dependent samples on a ranking scale \cite{42}. This method compares pairwise semantic relations between the samples by employing non-parametric sign tests, McNemar's tests, and the Wilcoxon test. To perform the test, an \(n\)-sized sample with paired data is assumed. In the null hypothesis, the difference between pairs with a symmetric distribution around zero is expressed as between zero and one. We calculate the value \(| X_{2,i} - X_{1,i} |\) for all pairs and identify significant differences while eliminating any zero differences. The dimensions of the new sample are designated as \( n_r \). The data is sorted by absolute value and ranked using the variable \(R_i\) \cite{43}. The Wilcoxon test is defined as
\begin{equation}
W = \sum_{i = 1}^{n_r} \left[ \mathrm{sign}(X_{2,i} - X_{1,i}) \times R_i \right].
\end{equation}
\textbf{Canonical Correlation Analysis (CCA):}
The CCA technique can identify linear relationships between two variables that have multiple dimensions. CCA accomplishes this by using complex labels to guide the selection of features based on their underlying semantics. Using two perspectives with the same conceptual element, CCA is able to retrieve the interpretation of the semantics. Let \((x, y)\) be a multivariate random vector. Assume that we have a set of observations, \( S = ((x_1, y_1), \ldots, (x_n, y_n)) \) of \((x, y)\) for this vector. We can represent the \(x\)-coordinates of these observations using \(S_x =  (x_1, \ldots, x_n)\) and the \(y\)-coordinates using \(S_y = (y_1, \ldots, y_n)\). To create a new coordinate for \(x\), we can choose a direction, \( w_x \) and project \(x\) in that direction, denoted \( x \rightarrow (w_x, X) \).
The function to be maximized is
\begin{equation}
\rho = \max_{w_x, w_y} \mathrm{corr}(S_x w_x, S_y w_y).
\end{equation}
The total covariance matrix \(C\) is a block matrix with two within-set covariance matrices \( C_{xx} \) and \( C_{yy} \), as well as two between-set covariance matrices \( C_{xy} = C_{yx}^{\top} \).
Therefore,
\begin{equation}
\rho = \max_{w_x, w_y} \frac{w_x^{\top} C_{xy} w_y}{\sqrt{w_x^{\top} C_{xx} w_x \,\, w_y^{\top} C_{yy} w_y}},
\end{equation}
which is the maximum canonical correlation obtained by optimizing over \(w_x\) and \(w_y\) \cite{44}.

\subsubsection{Multi-omics Analysis:}
Suppose we have \(n\) subjects, defined as \( y_i \ (i = 1, 2, \ldots, n)\) which are independently and identically distributed (IID). Each subject has covariates \(q-1\) denoted as
\( X_i = [X_{i1}, X_{i2}, \ldots, X_{iq}]^T \) and \(m\)-view datasets, \(M_i^{(1)}, \ldots, M_i^{(m)}\).
It is assumed that \( y_i\) follows a distribution in the exponential family with density,
\begin{equation}
f(y_i, \theta_i, \gamma) = \exp \left\{ \frac{y_i \theta_i - c_1(\theta_i)}{\gamma / w_i} + c_2(y_i,\gamma) \right\},
\end{equation}
where \(\theta_i\) and \(\gamma\) are the location and scale parameters, \( c_1(\cdot)\) and \( c_2(\cdot) \) are known functions, and \( W_i \) is a known weight. The mean and variance of \( y_i \) satisfy \( E(y_i) = c_1'(\theta_i) \) and \( \mathrm{Var}(y_i) = \gamma w_i = c_1''(\theta_i) \). In this generalized semiparametric model, we link the response variable \( y_i\) to a set of explanatory variables, which includes an intercept and \(m\)-view datasets:
\begin{equation}
g(y_i) = X_i^T \beta + f(M_i^{(1)}, \ldots, M_i^{(m)}).
\end{equation}

\begin{table*}[t]
  \centering
  \caption{Family and link functions of generalized linear models.}
  \label{tbl1}
  \begin{tabular*}{\textwidth}{@{\extracolsep{\fill}} llll @{}}
    \toprule
    \textbf{Error family} & \textbf{Link function} & \textbf{Inverse link} & \textbf{Typical use} \\
    \midrule
    Gaussian  & Identity, $g(y)=y$                          & $y=g^{-1}(y)$                    & Normally distributed data, $(-\infty,\,\infty)$ \\
    Gamma     & Inverse, $g(y)=1/y$                         & $g^{-1}(y)=1/y$                  & Positive continuous data, $(0,\,\infty)$ \\
    Binomial  & Logit, $g(y)=\log\!\left(\tfrac{y}{1-y}\right)$ & $g^{-1}(y)=\tfrac{e^{y}}{1+e^{y}}$ & Binary outcome data (0/1) \\
    Poisson   & Log, $g(y)=\log(y)$                         & $g^{-1}(y)=e^{y}$                & Count data with mean equal to variance \\
    \bottomrule
  \end{tabular*}
\end{table*}

The function \(g(\cdot)\) used in the model is a known monotonically increasing or decreasing link function, \(X_i\) is a \( q \times 1 \) vector of covariates including the intercept for the \(i\)-th subject, \(\beta\) is a \( q \times 1 \) vector of fixed effects, and \(f\) is an unknown function on the product domain,
\( M = M^{(1)} \otimes M^{(2)} \otimes \cdots \otimes M^{(m)} \) with \( M_i^{(l)} \in M^{(l)}, \ l=1,2,\ldots,m\).
We can decompose the function \(f\) using the ANOVA decomposition and represent it in a functional space (RKHS).\\
Let \( x_i \) denote the covariates \( (q-1) \), where \( X_{ij}, j=1, 2, \ldots, (q-1) \) is the quantity of the \( i \)-th subject. Also, let 
\[
M_i^{(1)} = [M_{i1}^{(1)}, \ldots, M_{id}^{(1)}], \quad
M_i^{(2)} = [M_{i1}^{(2)}, \ldots, M_{id}^{(2)}], \quad
M_i^{(3)} = [M_{i1}^{(3)}, \ldots, M_{id}^{(3)}],
\]
which correspond to genes with \(s\) SNP markers, \(d\) methylation profiles, and RNA-Seq profiles of the \( i \)-th subject. In this case:
\begin{equation}
\mathrm{logit}(P_i) = X_i^T \beta + f(M_i^{(1)}, M_i^{(2)}, M_i^{(3)}),
\end{equation}
where \(\mathrm{logit}(P_i) = \Pr(y_i = 1 \mid X_i, M_i^{(1)}, M_i^{(2)}, M_i^{(3)})\). Here, \( P_i \) represents the probability of \( i^{\text{th}} \) observation and \( \mathbf{p} = [p_1, p_2, \ldots, p_n]^T \). A linear mixed effects model can be used to model \(\mathbf{p}\) such that:
\begin{equation}
\begin{aligned}
\mathrm{logit}(P) &= X\beta + h_{M^{(1)}} + h_{M^{(2)}} + h_{M^{(3)}} \\
&\quad + h_{M^{(1)} \times M^{(2)}} + h_{M^{(1)} \times M^{(3)}} + h_{M^{(2)} \times M^{(3)}} \\
&\quad + h_{M^{(1)} \times M^{(2)} \times M^{(3)}}.
\end{aligned}
\end{equation}
Here, $\beta$ is a coefficient vector of fixed effects, 
$h_{M^{(1)}}, h_{M^{(2)}}, h_{M^{(3)}}, h_{M^{(1)} \times M^{(2)}}, h_{M^{(1)} \times M^{(3)}}, h_{M^{(2)} \times M^{(3)}}$, 
and $h_{M^{(1)} \times M^{(2)} \times M^{(3)}}$ are independent random effects with distributions 
$h_{M^{(1)}} \sim N(0, \tau^{(1)} K^{(1)})$, $\tau^{(1)} = \sigma^2 / \lambda^{(1)}$, 
$h_{M^{(2)}} \sim N(0, \tau^{(2)} K^{(2)})$, $\tau^{(2)} = \sigma^2 / \lambda^{(2)}$, 
$h_{M^{(3)}} \sim N(0, \tau^{(3)} K^{(3)})$, $\tau^{(3)} = \sigma^2 / \lambda^{(3)}$, 
$h_{M^{(1 \times 2)}} \sim N(0, \tau^{(1 \times 2)} K^{(1 \times 2)})$, 
$\tau^{(1 \times 2)} = \sigma^2 / \lambda^{(1 \times 2)}$, and so on. 
Estimation of variance components can be achieved by restricted maximum likelihood (ReML) \cite{45}.

\subsubsection{Testing marginal effects}
The kernel matrix for the respective view can be utilized to examine the marginal effect for every view data source. The null hypothesis of testing the marginal effect,
\[
H_0 : h_{M^{(l)}} = 0, \quad l = 1, 2, 3, 4, 5,
\]
is similar to measuring the variance components,
\[
H_0 : \tau^{(l)} = 0.
\]
The test statistic is
\[
S(\tau^{(l)}) = (y - X\beta)^T K^{(l)} (y - X\beta), \quad l = 1, 2, 3, 4, 5,
\]
which is the same as the sequence kernel association test (SKAT) \cite{46}.

\subsubsection{Testing interaction effect}
We can test interaction effects and higher-order interactions assuming no marginal effects, i.e., $\tau^{(l)} = 0$, $l = 1, \ldots, 5$. Testing the interaction effect
\[
H_0 : h_{M^{(l \times \xi)}} = 0, \quad l < \xi = 1, 2, 3, 4, 5,
\]
is comparable to evaluating the variance components,
\[
H_0 : \tau^{(l \times \xi)} = 0,
\]
with test statistic
\[
S(\tau^{(l \times \xi)}) = (y - X \beta)^T K^{(l \times \xi)} (y - X \beta), \quad \xi = 1, 2, 3, 4, 5.
\]
Similarly, we can test the effects of the third order interaction by assuming that all second order interactions are zero, i.e., $\tau^{(l)} = 0$ for $l = 1, \ldots, 5$, $\tau^{(l \times \xi)} = 0$, and
\[
H_0 : \tau^{(l \times \zeta \times \xi)} = 0,
\]
with statistic
\[
S(\tau^{(l \times \zeta \times \xi)}) = (y - X \beta)^T K^{(l \times \zeta \times \xi)} (y - X \beta), \quad l < \xi < \zeta = 1, 2, 3, 4, 5.
\]

\subsubsection{Statistical testing}
We discuss test statistics for the overall effect and various composite effects.

\subsubsection{Overall hypothesis testing}
With the KMR model, it is possible to test the overall effect using
\[
H_0 : h_{M^{(1)} \times M^{(2)} \times M^{(3)} \times M^{(4)} \times M^{(5)}} = 0,
\]
which is the same as assessing the variance component with
\[
H_0 : \tau^{1 \times 2 \times 3 \times 4 \times 5} = 0.
\]
The kernel matrices are not diagonally blocked and the variance component lies at the boundary under the null \cite{10,25,34}.

\begin{figure*}[ht!]
    \includegraphics[height=6cm,width=4cm]{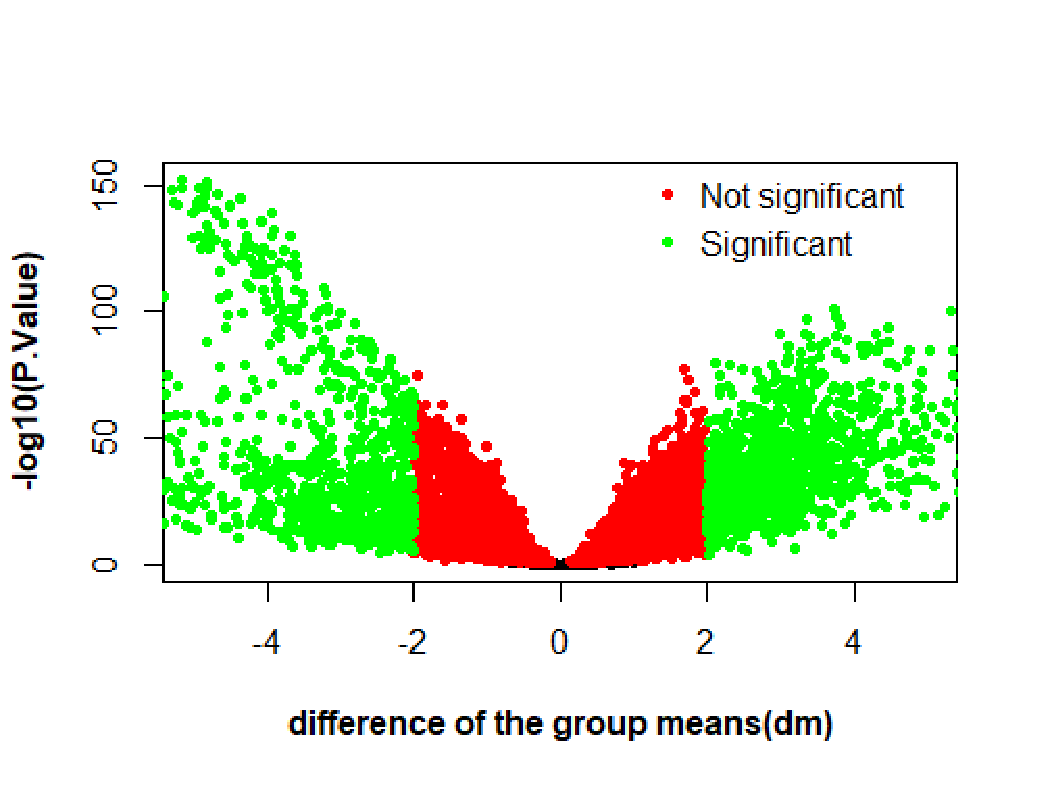}\hfill
    \includegraphics[height=6cm,width=4cm]{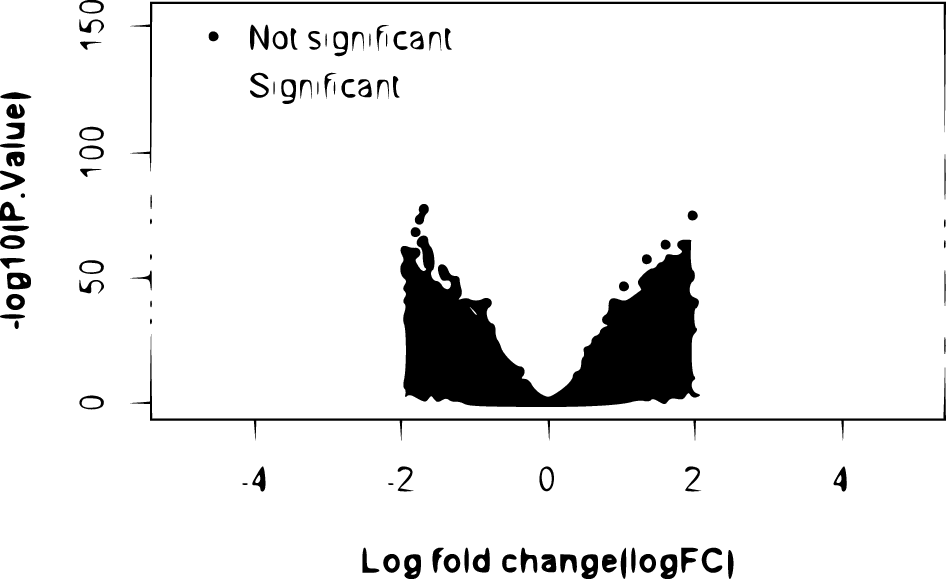}\hfill
    \includegraphics[height=6cm,width=4cm]{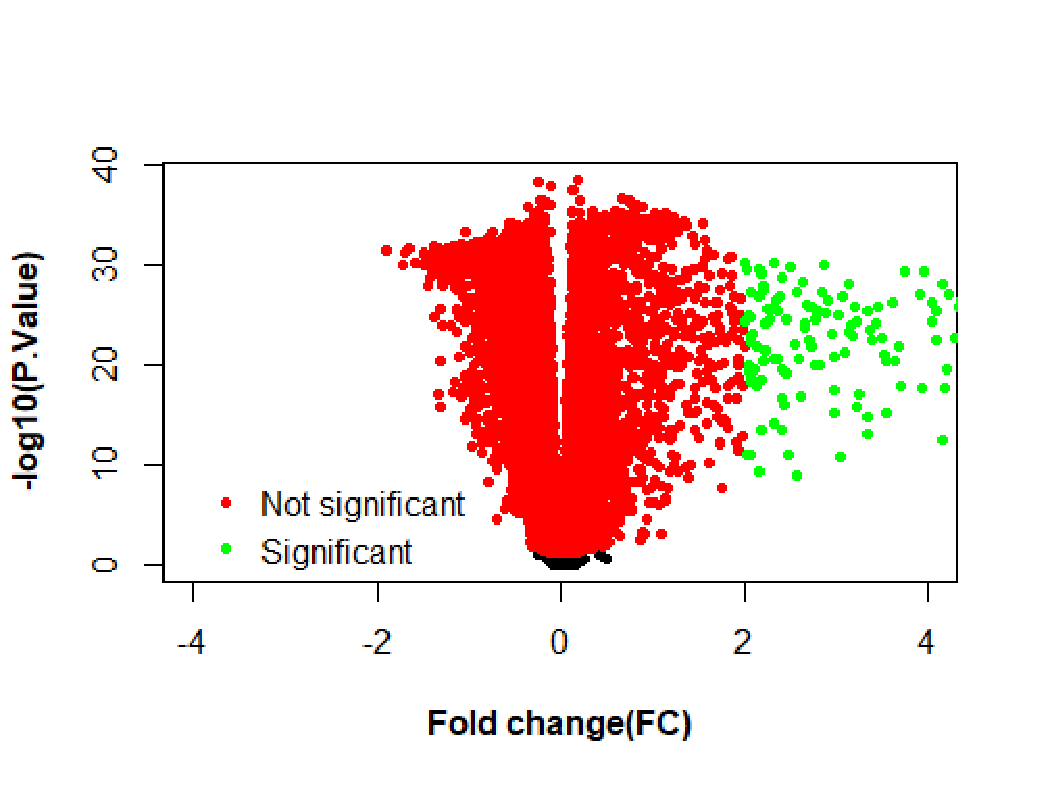}\hfill 
    \includegraphics[height=6cm,width=4cm]{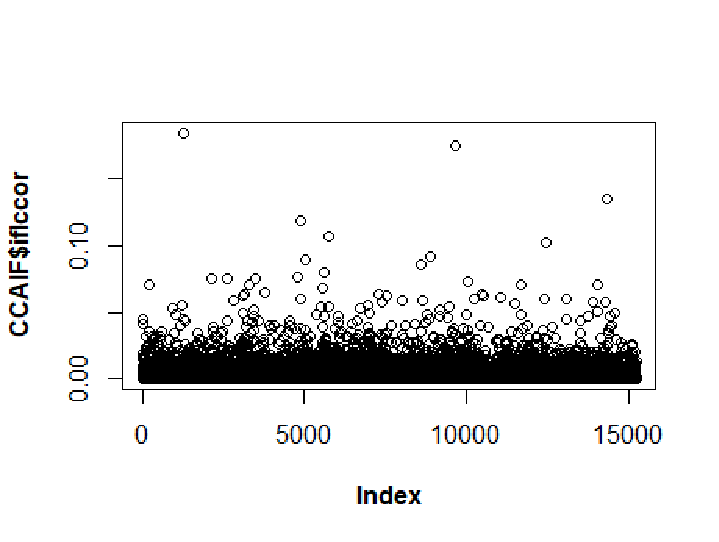}
    \caption{Differential gene expression analysis using (a) T-test, (b) LIMMA, (c) Wilcoxon, (d) CCA. See text for thresholds.}
    \label{FIG:3}
\end{figure*}

\subsubsection{Testing composite effects}
When lower-order effects show statistical significance, we may test for higher-order effects (composite testing). For the fifth-order composite effect,
\[
H_0 : h_{M^{(1)} \times M^{(2)} \times M^{(3)} \times M^{(4)} \times M^{(5)}} = 0,
\]
equivalently,
\[
H_0 : \tau^{1 \times 2 \times 3 \times 4 \times 5} = 0.
\]
Let \( \Sigma = \sigma^2 I + \tau^{(1)}K^{(1)} + \cdots + \tau^{(2 \times 3 \times 4 \times 5)}K^{(2 \times 3 \times 4 \times 5)} \), and all \(\tau\) and \(\sigma^2\) are model parameters under the null model. Define
\begin{equation}
S(\tilde{\theta}) = \frac{1}{2 \sigma_0^2} y^T B_I K^{(1 \times 2 \times 3 \times 4 \times 5)} y,
\end{equation}
where $\tilde{\theta} = (\sigma^2, \tau^{(1)}, \tau^{(2)}, \tau^{(3)}, \tau^{(1 \times 2)}, \tau^{(1 \times 3)}, \tau^{(2 \times 3)})$ and $B_I = \Sigma^{-1} - \Sigma^{-1}X(X^T\Sigma^{-1}X)^{-1}X^T\Sigma^{-1}$. The Satterthwaite method can approximate the distribution \cite{12}.

\subsection{Drug repurposing using molecular docking study}
We performed molecular docking of top-ranked proteins with drug agents to propose in-silico-validated candidate drugs for lung cancer. We collected 190 meta-drug agents from the literature [see \textcolor{blue}{Table~\ref{tbl5}}] to explore candidates. Protein 3D structures were downloaded from PDB \cite{47} and SWISS-MODEL \cite{48}. Drug 3D structures were downloaded from PubChem \cite{49}. Docking produced binding scores for each protein–drug pair \cite{50}. For protocol details see \cite{51}. Discovery Studio Visualizer 2019 \cite{52} was used to analyze docked complexes.

\begin{figure*}[ht!]
    \includegraphics[height=7cm,width=5.5cm]{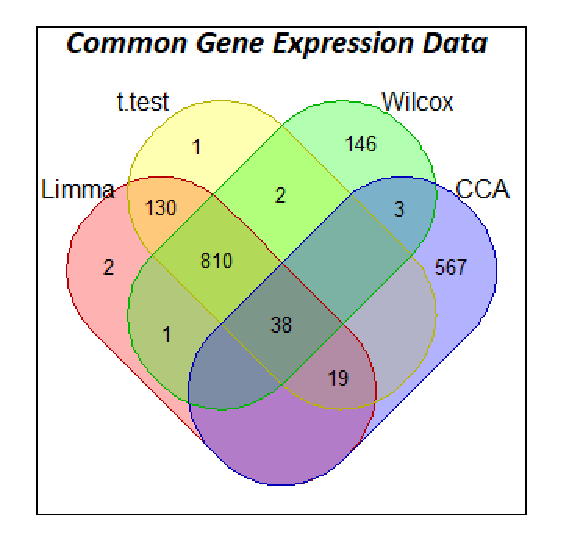}\hfill
    \includegraphics[height=7cm,width=5.5cm]{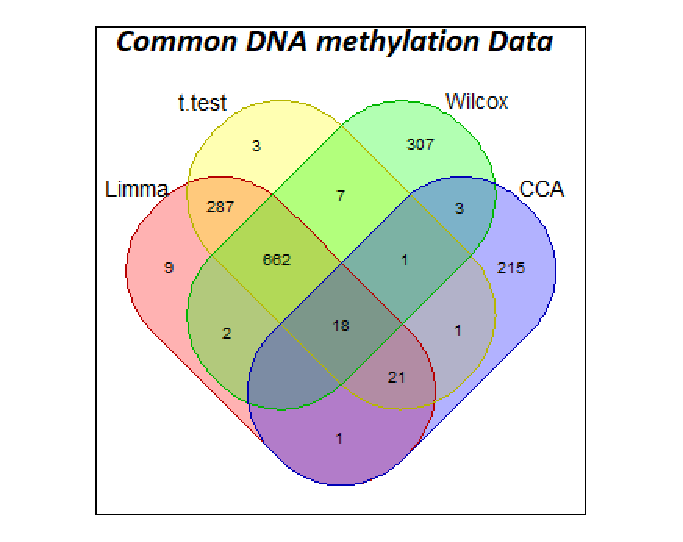}\hfill
    \includegraphics[height=7cm,width=5.5cm]{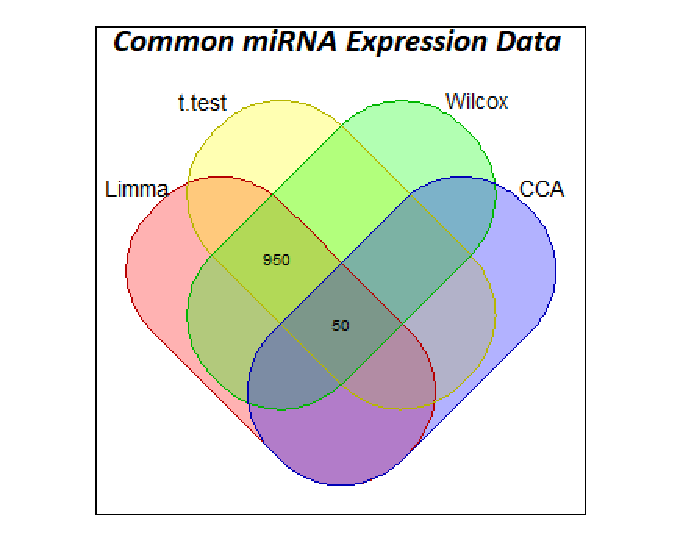}
    \caption{Venn diagrams for gene selection overlap across T-test, LIMMA, Wilcoxon, and CCA for (a) gene expression, (b) DNA methylation, (c) miRNA.}
    \label{FIG:6}
\end{figure*}

\section{Experimental  Analysis \& Results}\label{result}
We conducted experiments using three omics datasets from lung cancer studies. Our objective was to identify significant intersecting genes based on composite effects among omics. We used the Identity-By-State (IBS) kernel to analyze genetic similarity \cite{11}. Our goal was to identify genes harboring disease-associated variants by testing both common and rare variants and to discover potential drug targets. Our approach outperformed state-of-the-art group association tests in diverse scenarios \cite{19,39,40}. For interaction impact, Li and Cui (2012) used PCA-based approaches \cite{31}, while Alam et al.~(2018) treated each approach as a simple regression; we employed logistic regression \cite{11}.

\begin{table}[!htbp]
    \renewcommand{\arraystretch}{1.5}
    \caption{Selected significant genes using KMR (p-value threshold $= 0.00351$) for the top 9 triplets. OV: Overall effect; HOC: Higher-order composite effect.}
    \label{tbl2}
    \centering
        \scalebox{0.5}{
    \begin{tabular}{| c | c | c | c | c | c | c | c | c | c | c | c | c | }
        \hline
        Gene expression & miRNA expression & DNA methylation & $\sigma^2$ & $\tau^{(1)}$ & $\tau^{(2)}$ & $\tau^{(3)}$ & $\tau^{1 \times 2}$ & $\tau^{1 \times 3}$ & $\tau^{2 \times 3}$ & $\tau^{1 \times 2 \times 3}$ & OV & HOC \\
        \hline
        ABCA12 & SNAI1 & ABCC9 & 0.9963 & 0.0097 & 1.19E-06 & 0.0071 & 1.28E-07 & 0.0055 & 0.0087 & 0.01 & 0.4424 & 0.0008 \\
        \hline
        ABCA12 & PDGFRA & ABCC9 & 0.9986 & 0.0100 & 1.01E-06 & 0.0095 & 3.18E-06 & 0.0093 & 0.0101 & 0.0100 & 0.4670 & 0.0047 \\
        \hline
        ABCA12 & PDGFRB & SLCO1A2 & 1.0000 & 0.0100 & 0.0071 & 0.0100 & 3.24E-07 & 0.0100 & 0.0100 & 0.01 & 0.2948 & 0.0055 \\
        \hline
        F8 & ITGB1 & SLCO1A2 & 1.0000 & 0.0004 & 0.0009 & 0.0012 & 1.66E-08 & 0.0005 & 0.0010 & 0.0100 & 0.1190 & 0.0077 \\
        \hline
        HOXC13 & ITGB1 & ABCC9 & 1.0000 & 6.28E-09 & 3.72E-05 & 0.0009 & 0.0012 & 0.0009 & 0.0008 & 0.0100 & 0.2328 & 0.0057 \\
        \hline
        HOXC13 & MMP15 & SLCO1A2 & 1.0000 & 5.61E-08 & 0.0010 & 0.0010 & 0.0011 & 0.0009 & 0.0010 & 0.0100 & 0.1153 & 0.0068 \\
        \hline
        HOXC13 & SNAI1 & ASCL4 & 1.0000 & 0.0093 & 0.0039 & 1.65E-08 & 0.0102 & 0.0068 & 0.0068 & 0.0100 & 0.3306 & 0.0042 \\
        \hline
        PEBP4 & PTGFRN & SLCO1A2 & 1.0000 & 3.09E-08 & 0.0005 & 0.0009 & 0.0010 & 0.0006 & 0.0010 & 0.01 & 0.0035 & 0.0084 \\
        \hline
        SFTPC & ID1 & SLCO1A2 & 1.0000 & 0.0001 & 1.59E-06 & 9.31E-05 & 1.07E-06 & 9.85E-05 & 0.0001 & 0.0001 & 0.0671 & 0.0069 \\
        \hline
    \end{tabular}
    }
\end{table}

\subsection{Real data analysis}
In our real-world data analysis, we used the KMR method to examine three different omics datasets obtained from studies on lung cancer \cite{38}.

\subsection{Application to Lung Cancer study}
To apply the KMR method in the LUSC datasets, we treated each gene in the gene expression, DNA methylation and miRNA expression data as an individual evaluation unit. Due to high dimensionality, we reduced it by identifying differentially expressed genes using LIMMA, T-test, CCA, and Wilcoxon.

\begin{table}
    \caption{Number of genes identified as significant at various p-values using KMR.}\label{tbl3}
    \renewcommand{\arraystretch}{1.5}
    \setlength{\tabcolsep}{3pt}
    \centering
    \begin{tabular}{|p{1cm}|p{2cm}|p{2cm}|p{2cm}|}
    \hline
    \multicolumn{1}{|c|}{\textbf{P-values}} & \multicolumn{3}{c|}{\textbf{KMR}} \\
    \cline{2-4}
    & \textbf{Gene expression} & \textbf{miRNA expression} & \textbf{DNA methylation} \\
    \hline
    0.05 & 23  & 28 & 14\\
    \hline
    0.01 & 18  & 19 & 10\\
    \hline
    0.001 & 9 & 7  & 5\\
    \hline
    0.0001 & 2  & 3 & 3\\
    \hline
    \end{tabular}
\end{table}

\textcolor{blue}{Fig.~\ref{FIG:3}} and \textcolor{blue}{Figs.~1 and 2} (in supplementary) present differential analysis for gene expression, miRNA expression, and DNA methylation, respectively. For DEGs in \textcolor{blue}{Fig.~\ref{FIG:3}}, thresholds were absolute $dm>2$ and $p<0.05$. For miRNA and methylation (supplementary Figs.~1 and 2), DEGs used absolute fold change $>2$ and $p<0.05$.

\begin{figure}[ht!]
   \centering
   \includegraphics[height=15cm,width=15cm]{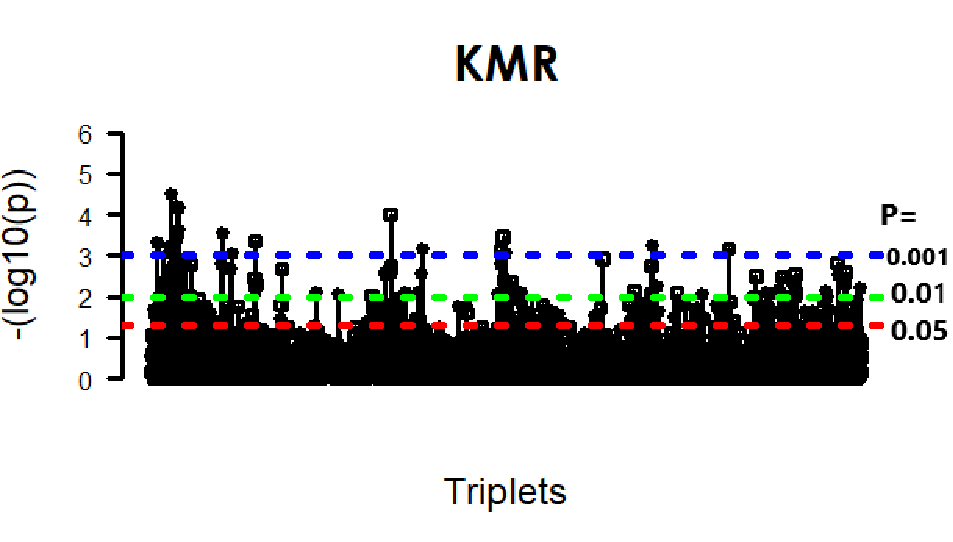}
   \caption{Manhattan plots of $-\log_{10}(p)$ versus triplets based on KMR overall tests.}
   \label{FIG:7}
\end{figure}

\begin{table*}[h]
    \renewcommand{\arraystretch}{3}
    \caption{The genes that were deemed significant by the KMR method at a p-value of 0.01. Only the genes highlighted by KMR were identified.}
    \label{tbl4}
    \setlength{\tabcolsep}{13pt}
        \scalebox{0.7}{
    \begin{tabular}{|c|p{13cm}|}
    \hline
    \multicolumn{1}{|c|}{\textbf{Method}} & \multicolumn{1}{c|}{\textbf{Genome}} \\
    
    \hline
     Genes &  ABCA12, ACOXL, C18orf56, CARD14, CD300C, CHRNA5, DQX1, F8, FCN1, FGF11, HOXC13, LRRC32, MUSTN1, PEBP4, SFTPC, SLC19A3, SORBS1, TNXB,  \\
    \hline
     DNA methylation &   HCFC1, ABCC9, SLCO1A2, ZIC1, PANCR, IPMK, HOXD3, GRM1, ASCL4,  \\
    \hline
     miRNA expression &  SNAI1, PDGFRA, PDGFRB, TMEM181, ANKRA2, PPP1R18, CD99, PTGFRN, GALNT7, SYNGR3, SMARCD2, GFPT2, PIP4K2A, ITGB1, AVEN, RTN4R ,TDG, MMP15, ID1 \\
    \hline
    \end{tabular}
    }
\end{table*}

\textcolor{blue}{Fig.~\ref{FIG:6}} shows the Venn diagrams for the three datasets. For gene expression, 1000 genes were exclusively selected by all four methods; for DNA methylation, 1000 genes were exclusively selected by all methods except CCA (260). For miRNA, CCA selected the fewest (50). Commonly selected by all methods were 38 (gene expression), 50 (miRNA), and 18 (methylation). We tested $34{,}200$ ($38 \times 50 \times 18$) triplets; \textcolor{blue}{Fig.~\ref{FIG:7}} shows $-\log_{10}(p)$ for all triplets. The overall test revealed 234 significant triplets at $p<0.05$; solid/dashed/dotted lines mark $0.05/0.01/0.001$.

\textcolor{blue}{Table~\ref{tbl2}} presents ReML estimates $\sigma^2,\tau^{(1)},\tau^{(2)},\tau^{(3)},\tau^{1\times2},\tau^{1\times3},\tau^{2\times3},\tau^{1\times2\times3}$ with corresponding $p$-values per triplet. At $p<0.00351$, we identified 9 significant triplets spanning five genes (ABCA12, F8, HOXC13, PEBP4, SFTPC), seven transcriptomes (SNAI1, PDGFRA, PDGFRB, ITGB1, MMP15, PTGFRN, ID1), and three epigenomes (ABCC9, SLCO1A2, ASCL4). \textcolor{blue}{Table~\ref{tbl3}} summarizes counts by threshold; \textcolor{blue}{Table~\ref{tbl4}} lists significant genes at $p=0.01$.\\

We generated gene-gene interaction networks with STRING (supplementary Fig.~3). Key metrics—nodes, edges, expected edges, average degree, clustering coefficient, and PPI enrichment $p$—were 44, 68, 43, 3.09, 0.498, and 0.000229, indicating high interconnectivity.

\begin{figure}[!htbp]
   \centering
   \includegraphics[height=10cm,width=15cm]{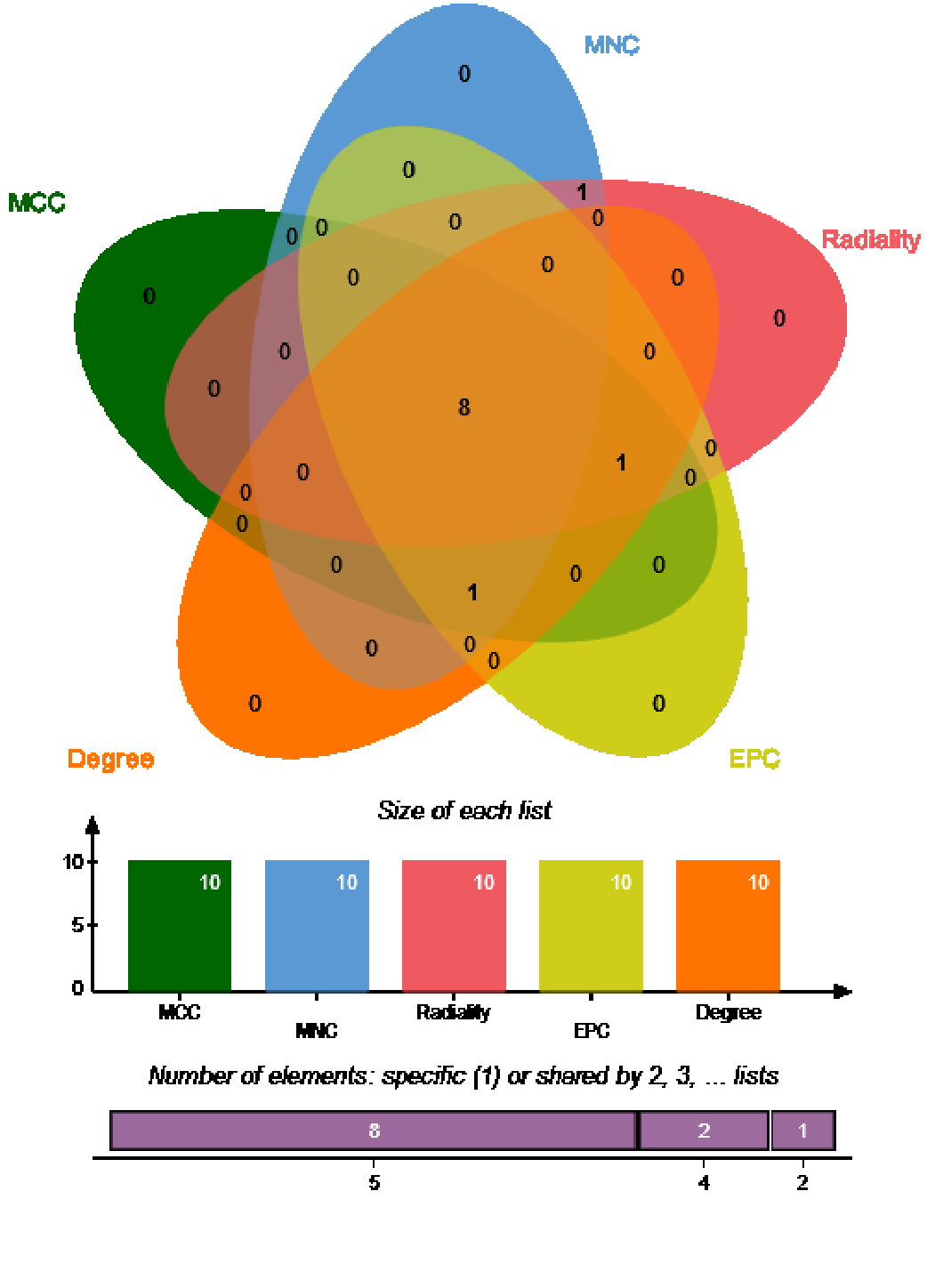}
   \caption{Venn diagram of hub genes across topological algorithms (MCC, MNC, Radiality, EPC, Degree).}
   \label{FIG:9}
\end{figure}

Cytoscape/CytoHubba identified top hub genes PDGFRA, ITGB1, SNAI1, FGF11, PDGFRB, ID1, TNXB, ZIC1 (Fig.~\ref{FIG:9}). We also evaluated classification precision using the identified features from KMR.

\begin{figure*}[ht!]
\centering
{\includegraphics[height=10cm,width=16cm]{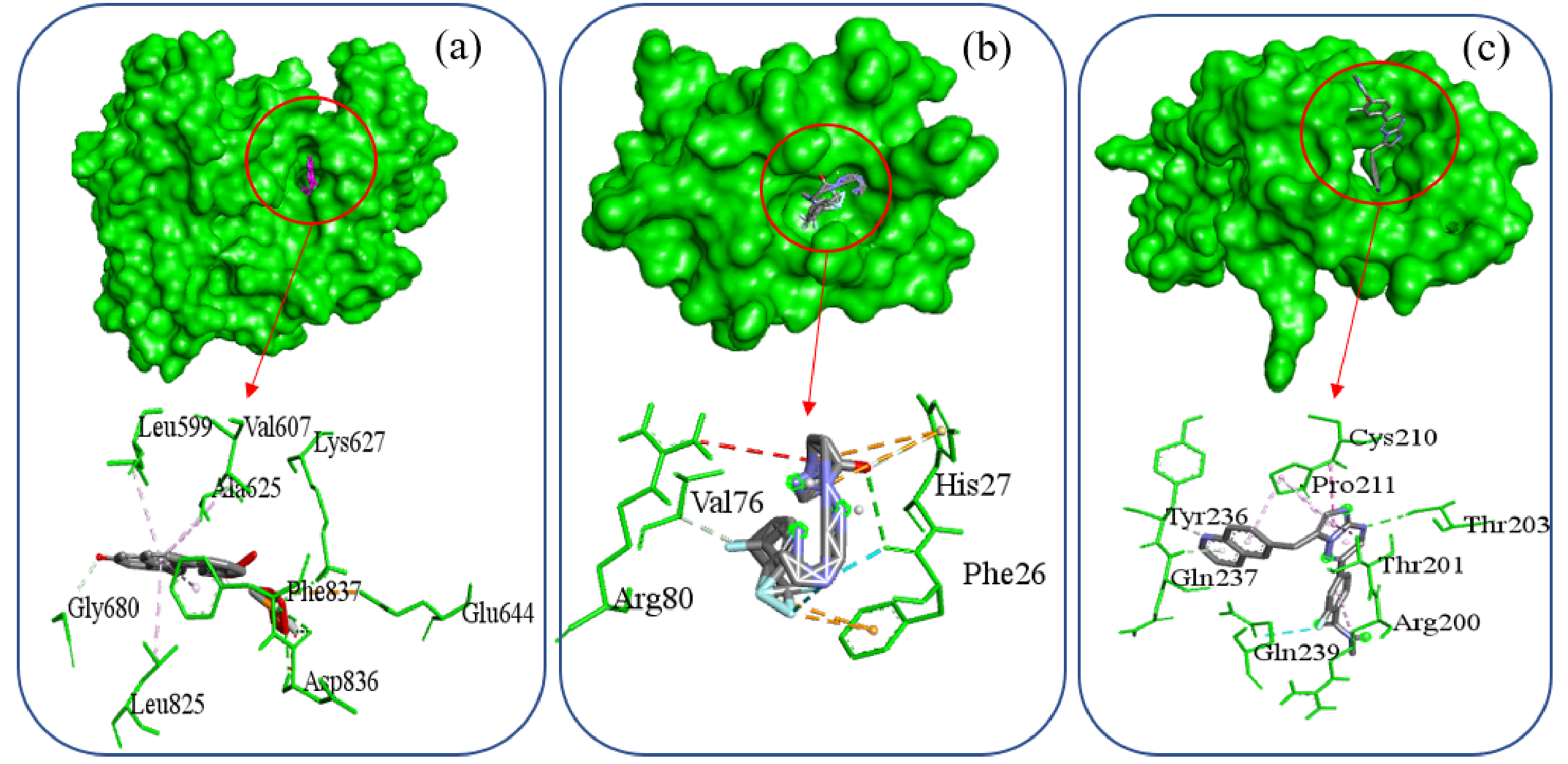}}
    \caption{Top-ranked three docked complexes and 2D chemical interactions.}
    \label{FIG:11}
\end{figure*}

\begin{table*}[t]
    \renewcommand{\arraystretch}{2}
    \caption{Interacting amino acids for the top three docked complexes with top three compounds.}
    \label{tbl5}
    \centering
    \begin{tabular}{|p{1cm}|p{1cm}|p{1cm}|p{3cm}|p{1cm}|p{1cm}|}
    \hline
    \textbf{Complex} & \textbf{Interacting residues} & \textbf{H-bond} & \textbf{Hydrophobic} & \textbf{Electrostatic} & \textbf{Halogen} \\ \hline
    PDGFRB--Selinexor & Glu644, Lys627, Asp836, Gly680, Leu599, Val607, Ala625, Leu825, Phe837 & Lys627, Asp836, Gly680 & Leu599, Val607, Ala625, Leu825, Phe837 & Glu644 & -- \\ \hline
    PDGFRA--Orapred & Phe26, Val76, His27, Arg80 & Phe26, Val76 & His27 & His27, Phe26 & Phe26 \\ \hline
    SNAI1--Capmatinib & Thr203, Tyr236, Gln239, Gln237, Arg200, Thr201, Cys210, Pro211 & Thr203, Tyr236, Gln237 & Arg200, Thr201, Cys210, Pro211 & -- & Gln239 \\ \hline
    \end{tabular}
\end{table*}

\subsection{Drug repurposing}
Drug repurposing identifies new indications for existing or investigational drugs \cite{53}. Molecular docking assesses binding affinities between ligands and targets \cite{54}. We considered eight proteins as targets and docked 190 meta-drug agents. PDB entries for PDGFRB, PDGFRA, SNAI1, TNXB, ITGB1 were 5grn, 1gq5, 3w5k, 2cum, 3g9w. ID1, FGF11, ZIC1 were modeled via SWISS-MODEL (UniProt P41134, Q6LA99, Q15915). The top three drugs (Selinexor, Orapred, Capmatinib) showed best aggregate binding (e.g., $-7.5$ kcal/mol). Complex interaction details are in Fig.~\ref{FIG:11} and Table~\ref{tbl5}. Independent studies support these candidates: Selinexor efficacy in KRASmut lung cancers \cite{55}; Orapred liposomal formulations show anti-angiogenic antitumor effects; Capmatinib is a selective MET inhibitor with robust responses \cite{56}.

\section{Conclusion}\label{conclusion}
We presented a KMR-based integrative analysis identifying higher-order composite effects across multi-omics in lung cancer. Using LIMMA, t-test, CCA, and Wilcoxon for feature preselection, and KMR for interaction testing, we found significant triplets and highlighted hub genes (PDGFRA, ITGB1, SNAI1, FGF11, PDGFRB, ID1, TNXB, ZIC1). Network analyses suggest these genes are highly interconnected. Drug repurposing via docking pointed to Selinexor, Orapred, and Capmatinib as promising candidates. Our results support KMR as a robust approach for multi-omics integration and target discovery, contingent on high-quality input data.

\bibliographystyle{IEEEtran}
\bibliography{refall}

\end{document}